\documentstyle{article}
\begin{document}

\title{Classical and quantum theories of spin}
\author{Fabi\'an H. Gaioli and Edgardo T. Garcia Alvarez \\
{\it Instituto de Astronom\'\i a y F\'\i sica del Espacio (IAFE), }\\
{\it \ C.C. 67, Suc. 28, 1428 Buenos Aires, Argentina }\\
{\it \ \& }\\
{\it \ Departamento de F\'\i sica, Facultad de Ciencias Exactas y Naturales, 
}\\
{\it Universidad de Buenos Aires, 1428 Buenos Aires, Argentina }}
\maketitle

\begin{abstract}
A great effort has been devoted to formulate a classical relativistic theory
of spin compatible with quantum relativistic wave equations. The main
difficulty in order to connect classical and quantum theories rests in
finding a parameter which plays the role of proper time at a purely quantum
level. We present a partial review of several proposals of classical and
quantum spin theories from the pioneer works of Thomas and Frenkel,
revisited in the classical BMT work, to the semiclassical model of Barut and
Zanghi [Phys. Rev. Lett. 52, 2009 (1984)]. We show that the last model can
be obtained from a semiclassical limit of the Feynman proper time
parametrization of the Dirac equation. At the quantum level we derive spin
precession equations in the Heisenberg picture. Analogies and differences
with respect to classical theories are discussed in detail.
\end{abstract}

\section{Introduction}

The interest in formulating classical theories to study the precession of
the spin lays in the fact that many problems, experimentally tested, could
be more simply derived from a consistent set of covariant classical
equations of motion in proper time. However at the quantum level the
standard formulation of relativistic wave equations did not explicitly
include the concept of proper time. Then some new ingredients are needed in
order to retrieve the covariant classical equations of motion starting from
quantum theory. In the semiclassical treatments of Rubinow and Keller \cite
{rk63} and Rafanelli and Schiller \cite{rs64} the proper time was introduced
through a WKB expansion of the Dirac equation. On the other hand many
quantum proper time derivatives were proposed in the literature \cite{a95a}.
These are heuristic proposals based on classical analogies, which go back to
the earlier works of Fock \cite{f37} and Beck \cite{b42}. Later on, in the
sixties, there was a growing interest for parametrized theories in an
invariant parameter (which looks like the proper time), known under several
names such as ``proper time'' formalisms or parametrized relativistic
quantum theories (PRQT). These formulations have their origin in the works
of St\"uckelberg \cite{s4142}, Feynman \cite{f48,f50,f51}, Nambu \cite{n50},
and Schwinger \cite{s51}. From the point of view of PRQT, ``proper time''
derivatives can be thought of as the derivatives with respect to the
evolution parameter in the Heisenberg picture.

In Section 2 we sketch the main lines of the classical theories of spin. For
this case we put special emphasis in the traditional point of view, as the
one developed by Frenkel \cite{f26} and Thomas \cite{t27} up to the revival
of such pioneer works carried out by Bargmann, Michel, and Telegdi \cite
{bmt59}. In Section 3 we revisit the most relevant quantum theories of
spinning systems and show how to reobtain the BMT equation in the classical
limit in different ways. In Section 4 we focus our attention on the theory
associated with the proper time parametrization of the Dirac equation
originally proposed by Feynman \cite{f48} and we discuss about the necessity
to formulate an adequate framework out of the mass shell. We show the way to
connect this formalism with the results of Section 3. Finally in Section 5,
using the formalism of Section 4, we rederive in the semiclassical limit,
via relativistic coherent states, the formulation of Barut and Zanghi \cite
{bz84} of the classical spinning system.

\section{Classical theories}

Classical theories of spinning particles were formulated as a generalization
to any inertial frame of the classical spin precession in the presence of
external electromagnetic fields (${\bf E},{\bf B}$) in the rest frame,
namely ($c=1$)

\begin{equation}
\frac{d{\bf S}}{dt}={\bf \mu }\times {\bf B},  \label{uno}
\end{equation}
where ${\bf \mu }$ is the magnetic moment of the electron, which is assumed
to be proportional to the spin ${\bf S}$, ${\bf \mu }=\frac{ge}{2m}{\bf S}$,
being $m$ the mass, $e$ the charge, and $g$ the gyromagnetic factor of the
electron. It is also assumed that the electric moment is zero in the rest
frame.

There are two natural covariant generalizations to any frame of the spin $%
{\bf S}$ and consequently of Eq. (\ref{uno}). We can consider an axial
four-vector $S_\alpha $ or an antisymmetric second-order tensor $S_{\alpha
\beta }$, with only three independent components, such that $S_0=0$ and
equivalently $S_{0i}=0$ in the rest frame. In a covariant way these
conditions read \cite{t27,m37,wr47,p51}

\begin{equation}
S_\alpha u^\alpha =0;\hspace{1.0in}S_{\alpha \beta }u^\beta =0,  \label{dos}
\end{equation}
where $u^\alpha $ is the four-velocity $u^\alpha =\frac{dx^\alpha }{ds}$ (a
unit time-like vector $u^\alpha u_\alpha =1$). These objects are then
related by\footnote{%
The conventions are $\eta _{\mu \nu }={\rm diag}(+1,-1,-1,-1)$ for the
Minkowski metric tensor, $\varepsilon ^{0123}=1$ for the Levi- Civita
tensor, and $\gamma ^5=\gamma ^0\gamma ^1\gamma ^2\gamma ^3$.}

\begin{equation}
S^\alpha =\frac 12\varepsilon ^{\alpha \beta \mu \nu }u_\beta S_{\mu \nu };%
\hspace{1.0in}S_{\alpha \beta }=\varepsilon _{\alpha \beta \mu \nu }u^\mu
S^\nu .  \label{tres}
\end{equation}

Frenkel \cite{f26} and Thomas \cite{t27} considered the most general
equation for the proper time variation of the spin vector or tensor --this
includes non-minimal couplings between matter and external electromagnetic
fields-- compatible with Eq. (\ref{uno}), by taking into account all the
relativistic invariants one can construct, assuming that the precession
equation is linear in the spin ($S_\alpha $ or $S_{\alpha \beta }$) and in
the electromagnetic fields ($F_{\mu \nu }=\partial _\mu A_\nu -\partial _\nu
A_\mu $). Considering also the Lorentz force law\footnote{%
At this level we are neglecting gradients in the fields, e.g. ${\bf \nabla }%
\left( {\bf \mu }\cdot {\bf B}\right) $, but however retaining non-minimal
couplings at the level of the spin precession equation. Neglecting
consistently non-minimal terms Kramers \cite{k34} was able to demonstrate
that the gyromagnetic factor is $g=2$. This result is commonly believed to
be explained only by the Dirac equation.}

\begin{equation}
\frac{d\pi _\mu }{ds}=eF_{\mu \nu }u^\nu ,  \label{cuatro}
\end{equation}
where $\pi _\mu =mu_\mu $, the spin precession equations result

\begin{equation}
\frac{dS_\alpha }{ds}=\frac em\left[ \frac g2F_\alpha ^{\ \beta
}S_\beta +\left( \frac g2-1\right) u_\alpha \left( S_\lambda F^{\lambda \nu
}u_\nu \right) \right] ,  \label{cinco}
\end{equation}

\begin{equation}
\frac{dS_{\alpha \beta }}{ds}=\frac e{2m}\left( 1-\frac g2\right) \left(
S_{\beta \nu }u_\alpha -S_{\alpha \nu }u_\beta \right) F_\lambda ^{\ %
\nu }u^\lambda +\frac{ge}{2m}\left( S_{\alpha \lambda }F_\beta ^{\ %
\lambda }-S_{\beta \lambda }F_\alpha ^{\ \lambda }\right) .
\label{seis}
\end{equation}
Equation (\ref{cinco}) is the celebrated BMT equation, which was
rediscovered by Bargmann, Michel, and Telegdi \cite{bmt59}.

\section{Quantum theories}

\subsection{Semiclassical proper time derivative}

Rubinow and Keller \cite{rk63} continued and old attempt of Pauli \cite{p32}
to solve the Dirac equation for a particle in an electromagnetic field
minimally coupled, $\pi _\mu \equiv p_\mu -eA_\mu ,$ $p_\mu =i\partial _\mu $
($\hbar =1$),

\begin{equation}
\left( \gamma ^\mu \pi _\mu -m\right) \psi =0,  \label{siete}
\end{equation}
using the WKB method, proposing a solution of the form

\begin{equation}
\psi _{{\rm WKB}}=\psi _0e^{-iS},  \label{ocho}
\end{equation}
where $S$ is an scalar function and $\psi _0$ a spinor. They were able to
obtain the appropriate solution $\psi _{{\rm WKB}}$ also for the modified
Dirac equation of a particle with an anomalous magnetic moment (by adding to
Eq. (\ref{siete}) a Pauli term, $\left( \frac g2-1\right) \frac e{2m}F^{\mu
\nu }S_{\mu \nu }$, where $\frac g2-1$ is the anomalous gyromagnetic factor,
and $S_{\mu \nu }\equiv \frac 12\sigma _{\mu \nu }\equiv \frac i4\left[
\gamma _\mu ,\gamma _\nu \right] $). For any variable $a=a(x^\mu ,\gamma
^\mu )$ it can be associated a density ${\cal A}(x^\mu )$, such that

\begin{equation}
{\cal A}\equiv \overline{\psi }_{{\rm WKB}}a\psi _{{\rm WKB}},  \label{nueve}
\end{equation}
where $\overline{\psi }\equiv \psi ^{\dagger }\gamma ^0$ is the Dirac
adjoint. As particular cases of Eq. (\ref{nueve}) we have \cite{ga98}

\begin{equation}
{\cal U}_\mu =\frac 1m\left( \partial _\mu S-eA_\mu \right) =\overline{\psi }%
_{{\rm WKB}}\gamma _\mu \psi _{{\rm WKB}}  \label{diez}
\end{equation}
and

\begin{equation}
{\cal S}^\mu =\overline{\psi }_{{\rm WKB}}\frac i2\gamma ^5\gamma ^\mu \psi
_{{\rm WKB}},\hspace{1.0in}{\cal S}^{\mu \nu }=\overline{\psi }_{{\rm WKB}%
}S^{\mu \nu }\psi _{{\rm WKB}},  \label{once}
\end{equation}
which are the semiclassical densities corresponding to the Michel and
Wightman \cite{mw55} polarization four-vector $t^\mu \equiv i\gamma ^5\gamma
^\mu $ and to the spin tensor $S^{\mu \nu }$, which satisfy relations
analogous to those of Eq. (\ref{tres}). In order to obtain manifestly
covariant equations of motion Rubinow and Keller \cite{rk63} introduced the
concept of proper time derivative in this semiclassical approximation, for
which the notion of trajectory retrieves its meaning, through

\begin{equation}
\frac{d{\cal A}}{ds}=\frac{\partial {\cal A}}{\partial x^\mu }\frac{dx^\mu }{%
ds}={\cal U}^\mu \frac{\partial {\cal A}}{\partial x^\mu },  \label{doce}
\end{equation}
and derived the BMT equation for the spin vector (\ref{once}).

Rafanelli and Schiller \cite{rs64}, based on Fock's \cite{f37} work were
also able to reobtain the BMT equation (\ref{cinco}) in an easier way,
applying the WKB method to the squared Dirac equation. Following these ideas
we have shown \cite{ga98} a deeper connection with Fock's work, {\it i.e.}

\begin{equation}
\frac{d{\cal A}}{ds}=\overline{\psi }_{{\rm WKB}}\left( \frac{da}{ds}\right)
_{{\rm Fock}}\psi _{{\rm WKB}},  \label{trece}
\end{equation}
where the Fock proper time derivative will be defined in Section 3.2. As we
will see the BMT equation will be obtained as a particular case of (\ref
{trece}).

\subsection{Quantum proper time derivatives}

At a purely quantum level it is necessary to introduce the notion of proper
time derivative in order to write explicitly covariant equations of motion
in this parameter. The first in introducing the concept of proper time was
Fock \cite{f37}, through a parametrization which corresponds, in the
Heisenberg picture, to a proper time derivative

\begin{equation}
\left( \frac{da}{ds}\right) _{{\rm Fock}}\equiv -\frac i{2m}\left[
H^2,a\right] ,  \label{catorce}
\end{equation}
for any dynamical variable $a$, where $H\equiv \gamma ^\mu \pi _\mu $. For
example, for $a=S_{\mu \nu }$ we have

\begin{equation}
\left( \frac{dS_{\alpha \beta }}{ds}\right) _{{\rm Fock}}=\frac em\left(
S_{\alpha \lambda }F_\beta ^{\ \lambda }-S_{\beta \lambda }F_\alpha ^{%
\ \lambda }\right) ,  \label{quince}
\end{equation}
which is the analogous of Eq. (\ref{seis}) for $g=2$.

Some years later Beck \cite{b42} proposed a first-order proper time
derivative

\begin{equation}
\left( \frac{da}{ds}\right) _{{\rm Beck}}\equiv -i\left[ H,a\right] .
\label{dieciseis}
\end{equation}
Applying (\ref{dieciseis}) to $S_{\alpha \beta }$ we obtain \cite{a91}

\begin{equation}
\left( \frac{dS_{\alpha \beta }}{ds}\right) _{{\rm Beck}}=\pi _\alpha \gamma
_\beta -\pi _\beta \gamma _\alpha ,  \label{dieciocho}
\end{equation}
where\footnote{%
This relation is the covariant generalization of $\frac{d{\bf x}}{dt}={\bf %
\alpha }$ in Dirac's theory, which originated the localization problem.
Compare Eq. (\ref{diecinueve}) with Eq. (\ref{diez}).}

\begin{equation}
\gamma ^\mu =\left( \frac{dx^\mu }{ds}\right) _{{\rm Beck}}.
\label{diecinueve}
\end{equation}
Equation (\ref{dieciocho}) has not classical analogy yet. We will come back
on this relation in Section 4, where we show that by eliminating the
interference between positive and negative mass states (covariant {\it %
Zitterbewegung}) we retrieve the Frenkel-Thomas equation.\footnote{%
We are considering only the minimal coupling case. The generalization of
these results with the inclusion of a Pauli term into the covariant
Hamiltonian $H$ is straightforward.}

The last derivations of quantum equations of spin motion come from heuristic
proper time derivatives but not from first principles. They are formal
relations assumed to be valid on the mass shell. However there is no reason
for justifying Eq. (\ref{diecinueve}) which uses that

\begin{equation}
\lbrack x_\mu ,p_\nu ]=-i\eta _{\mu \nu },  \label{veinte}
\end{equation}
since at the classical level the covariant canonical Poisson brackets $%
\{x_\mu ,p_\nu \}=\eta _{\mu \nu }$ are incompatible with the constraint $%
\pi ^\mu \pi _\mu =m^2$.\footnote{%
It is easy to see that $0=\{x_\mu ,m^2\}=\{x_\mu ,\pi ^\nu \pi _\nu \}=2\pi
_\mu $, which leads to an absurd.} Then we need to extend our formalism
beyond the mass shell condition. This is the goal of Section 4.

\section{Feynman parametrization}

We are looking for a first quantized formalism which provides a consistent
unification of relativity and quantum mechanics. The main requirements we
need to take into account are \cite{a95a}:

\begin{quotation}
{\small $\bullet $ Dirac's theory must be somehow included. }

{\small $\bullet $ The framework must not be restricted to the mass shell. }

{\small $\bullet $ The equations of motion must be analogous to those of the
classical theory in which the evolution parameter is the proper time. }
\end{quotation}

\noindent The first and third conditions are the Bohr correspondence
principle, while the second one means that we have to enlarge the Poincar\'e
algebra to be able to include a four-vector position operator $x^\mu $
(which localizes the charge) that satisfies the canonical commutation
relations (\ref{veinte}). We have developed \cite{a92,a95b,gag96,gag97,gag98}
a consistent theory based on the Feynman parametrization of the Dirac
equation

\begin{equation}
-i\frac d{ds}\left| \psi \right\rangle =H\left| \psi \right\rangle .
\label{veintiuno}
\end{equation}
Here we only want to briefly discuss the three points raised above. Equation
(\ref{veintiuno}) is a Sch\"ordinger-like equation where the evolution
parameter can be identified with the proper time. Dirac equation is
recovered as an eigenvalue equation for the mass operator $H$,

\begin{equation}
H\left| \psi _m\right\rangle =m\left| \psi _m\right\rangle ,
\label{veintidos}
\end{equation}
where $\left| \psi _m\right\rangle $ are stationary states in $s$. So, in
general, an arbitrary state in this theory does not have a definite mass.
These facts explain the first two points. Now let us focus our attention on
the third one. The Beck derivative (\ref{dieciseis}) becomes the equation of
motion for the dynamical variables in the Heisenberg picture. We now
consider semiclassical relativistic coherent states of the form $\left| \psi
\right\rangle _c=u\left| \varphi \right\rangle _c$, where $u$ is a constant
spinor on the positive mass shell and $\left| \varphi \right\rangle _c$ is a
relativistic coherent state for the orbital part, such that 
\begin{equation}
\Lambda _{+}\left| \psi \right\rangle _c=\left[ 1+O(\hbar )\right] \left|
\psi \right\rangle _c,  \label{condic}
\end{equation}
where $\Lambda _{+}\equiv \frac 12\left( I+\frac{\gamma ^\mu p_\mu }m\right) 
$ is the projector on positive mass states. Taking mean values with these
semiclassical states Eq. (\ref{dieciseis}) becomes \cite{a95b}

\begin{equation}
\left\langle \frac{da}{ds}\right\rangle _c=-\frac i{2m_c}\left\langle \left[
\pi ^\mu \pi _\mu -eF^{\mu \nu }S_{\mu \nu },a\right] \right\rangle _c,
\label{veintitres}
\end{equation}
where $m_c\equiv \left\langle \left( \pi ^\mu \pi _\mu \right) ^{\frac
12}\right\rangle _c$ plays the role of a classical variable mass. Equation (%
\ref{veintitres}) resembles Fock's derivative. The important relation (\ref
{veintitres}) allows us to retrieve the classical equations of motion in the
limit $\hbar \rightarrow 0$, {\it e.g.} it allows us to recover the BMT
equation for $g=2$ [using the result of (\ref{quince}) into Eq. (\ref
{veintitres})]. This formalism also solves the problems of relativistic wave
equations --which were the cause of the reformulation given by quantum field
theory-- consistently incorporating into a one-charge theory the presence of
particles and antiparticles.\footnote{%
Feynman \cite{f49} has pointed out: ``The various creation and annihilation
operators in the conventional electron field view are required because the
number of particles is not conserved, {\it i.e.}, pairs must be created or
destroyed. On the other hand charge is conserved which suggests that if we
follow the charge, not the particle, the results can be simplified.''}

In Refs. \cite{gag96,gag97} we have integrated the Heisenberg equations of
motion for $\pi _\mu $ and $\gamma _\mu $, {\it i.e.}

\begin{equation}
\frac{d\gamma _\mu }{ds}=4S_{\mu \nu }\pi ^\nu ,  \label{veinticuatro}
\end{equation}

\begin{equation}
\frac{d\pi _\mu }{ds}=eF_\mu ^{\ \nu }\gamma _\nu ,
\label{veinticinco}
\end{equation}
in order to study the covariant {\it Zitterbewegung} in external
electromagnetic fields (see also Ref. \cite{bt85b}). It has permitted us to
describe particle creation in a constant external electric field. In such a
case we have obtained a pictorial representation of the corresponding
Feynman diagram as a {\it zig-zag} motion of the charge in space-time.

Summarizing, we have seen that it can be formulated a consistent (off-shell)
framework from which to recover the classical covariant (on-shell) equations
of motion in the semiclassical limit. In Section 5 we are going to study the
semiclassical limit of the Feynman parametrization keeping the covariant 
{\it Zitterbewegung} without projecting on the positive mass shell.

\section{Barut-Zanghi model for spin}

The Feynman parametrization given in Eq. (\ref{veintiuno}) can be derived
from the action

\begin{equation}
I=\left[ -i\overline{\psi }(x^\mu ,s)\frac{\partial \psi (x^\mu ,s)}{%
\partial s}-\overline{\psi }(x^\mu ,s)\gamma ^\mu \pi _\mu \psi (x^\mu
,s)\right] d^4xds.  \label{veintiseis}
\end{equation}
Performing a semiclassical limit as the one considered in Eq. (\ref
{veintitres}), $\psi (x^\mu )=z\varphi _c(x^\mu )$, where $z$ is now an
arbitrary constant spinor,\footnote{%
In this case the spinor $z$, in general, does not satisfy condition (\ref
{condic}).} we have \cite{ga98}

\begin{equation}
I_c=\left[ -i\overline{z}\frac{dz}{ds}+\epsilon \frac{d\left\langle x^\mu
\right\rangle _c}{ds}\left\langle p_\mu \right\rangle _c-\overline{z}\gamma
^\mu z\left\langle \pi _\mu \right\rangle _c\right] ds,  \label{veintisiete}
\end{equation}
where $\epsilon \equiv \overline{z}z=\pm 1$ takes into account particle and
antiparticle states, according to the St\"uckelberg interpretation.\footnote{%
A discussion of the importance of the factor $\epsilon $ in derivating the
Dirac equation, starting from the St\"uckelberg interpretation, can be seen
in Ref. \cite{gga95}.} From now on we will drop the averages $\left\langle 
\ \right\rangle _c$ interpreting all orbital variables as mean values
in semiclassical states. The semiclassical action (\ref{veintisiete}) is the
action proposed by Proca \cite{p54} and Barut and Zanghi \cite{bz84} as a
classical theory for a spinning system that undergoes a real {\it %
Zitterbewegung}.\footnote{%
The quantization of this theory was performed by Barut and Pavsic \cite{bp87}
and Barut and Un\"al \cite{bu93}.} Nevertheless in their formulation $%
\epsilon $ does not appear so that their action only describes particle
motions.\footnote{%
Notice that for the temporal component of the first of Eqs. (\ref
{veintinueve}) we have $\frac{dx^0}{ds}=\epsilon z^{\dagger }z$. As $%
z^{\dagger }z$ is always positive, according to the St\"uckelberg \cite
{s4142} interpretation, it is impossible to have antiparticles unless $%
\epsilon =-1$.}

The Hamilton equations derived from the action (\ref{veintisiete}) in the
extended classical phase-space of coordinates $(z,\overline{z},x^\mu ,p_\mu
) $ are

\begin{equation}
\frac{dz}{ds}=i\gamma ^\mu \pi _\mu z,\hspace{1.0in}\frac{d\overline{z}}{ds}%
=-i\overline{z}\gamma ^\mu \pi _\mu ,  \label{veintiocho}
\end{equation}
\begin{equation}
\frac{dx^\mu }{ds}=\epsilon \overline{z}\gamma ^\mu z,\hspace{1.0in}\frac{%
d\pi _\mu }{ds}=\epsilon eF^{\mu \nu }\overline{z}\gamma ^\nu z.
\label{veintinueve}
\end{equation}
From these equations we have that ${\cal H}\equiv u^\mu \pi _\mu $ is a
constant of motion (the analogue of Feynman's Hamiltonian $\gamma ^\mu \pi
_\mu $), where $u^\mu \equiv \epsilon \overline{z}\gamma ^\mu z$ [{\it cf.}
Eq. (\ref{diecinueve})]. Free particle solutions are

\begin{eqnarray}
u^\mu (s) &=&p^\mu \frac{{\cal H}}{p^2}+\left[ u^\mu (0)-p^\mu \frac{{\cal H}%
}{p^2}\right] \cos (2ps)+\frac 1{2p}\frac{du^\mu (0)}{ds}\sin (2ps), 
\nonumber \\
&&  \label{treinta} \\
p_\mu &=&{\rm const},  \nonumber
\end{eqnarray}
where $p\equiv \sqrt{p^\mu p_\mu }$ can be identified with the positive mass
of the particle ($p$ enters here as a constant of the motion). Equations (%
\ref{treinta}) show the classical analog of the phenomenon of {\it %
Zitterbewegung} of the charge [compare them with the operator solution
obtained by Barut and Thacker \cite{bt85a} and in Refs. \cite{gag96,gag97}].
The four-velocity has the smooth term $p^\mu \frac{{\cal H}}{p^2}$ --as it
is expected for the center of mass of particles-- which is the term free of
interferences between positive and negative masses, {\it i.e.} $p^\mu \frac{%
{\cal H}}{p^2}=\overline{z}\left( \Lambda _{+}\gamma ^\mu \Lambda
_{+}+\Lambda _{-}\gamma ^\mu \Lambda _{-}\right) z$. The other contribution
to the velocity is an oscillatory motion with the characteristic frequency $%
\omega =2p$, which is a covariant classical analog of the {\it Zitterbewegung%
}. The spin results to be the orbital angular momentum of the {\it %
Zitterbewegung} \cite{h52,bb81,bt85a}.

As Barut and Zanghi \cite{bz84} have pointed out their formulation has a
more natural form in a five-dimensional space-time with metric signature $%
(+,-,-,-,-)$. In Refs. \cite{gag96,gag97,gag98} we have shown that the
Feynman parametrization (\ref{veintiuno}) naturally arises as a ``massless''
Dirac equation in this manifold.

Instead of the variables $z$ and $\overline{z}$ we can work in terms of the
spin variables. Then we can obtain \cite{bz84,bc93} a new set of dynamical
equations equivalent to (\ref{veintiocho}) and (\ref{veintinueve}):

\begin{equation}
\frac{dx^\mu }{ds}=u^\mu ,  \label{treintayuno}
\end{equation}
\begin{equation}
\frac{du_\mu }{ds}=4S_{\mu \nu }\pi ^\nu ,  \label{treintaydos}
\end{equation}
\begin{equation}
\frac{d\pi _\mu }{ds}=eF_{\mu \nu }u^\nu ,  \label{treintaytres}
\end{equation}
\begin{equation}
\frac{dS_{\alpha \beta }}{ds}=\pi _\alpha u_\beta -\pi _\beta u_\alpha ,
\label{treintaycuatro}
\end{equation}
for the new set of dynamical variables $(x^\mu ,u^\mu ,p_\mu ,S_{\mu \nu })$%
, where now $S_{\mu \nu }\equiv \frac 14i\epsilon \overline{z}[\gamma _\mu
,\gamma _\nu ]z$. Equation (\ref{treintayuno}) must be compared with (\ref
{diecinueve}) and (\ref{diez}), while Eqs (\ref{treintaytres}) and (\ref
{treintaycuatro}) are respectively the Lorentz force law and the spin
precession equation for the spin tensor in the case of minimal coupling [%
{\it cf.} Eq. (\ref{dieciocho})].\footnote{%
The classical relativistic spinning particle with anomalous magnetic moment
was considered by Barut and Cruz \cite{bc93}, who arrived to the BMT
equation after averaging in $s$.} Equations (\ref{treintaydos}) and (\ref
{treintaytres}) are the analogous of Eqs. (\ref{veinticuatro}) and (\ref
{veinticinco}) respectively.\footnote{%
The analogies between the equations of motion obtained from the Beck
derivative and those obtained from the Barut and Zanghi classical
formulation were also remarked by Barut and Un\"al \cite{bu93}.} In this
semiclassical formulation conditions (\ref{dos}) and $u^\alpha u_\alpha =1$
are only satisfied after taking the projection on the positive mass
subspace. This eliminates the covariant {\it Zitterbewegung} and it is
essentially equivalent to average in $s$.\footnote{%
See, for example, Eq. (\ref{treinta}) in which, after averaging in $s$, only 
$p^\mu \frac{{\cal H}}{p^2}$ remains.} In this case, in the same way as it
happened for Eq. (\ref{veintitres}), we reobtain the form of the classical
equations of Frenkel and Thomas.

We have seen that the modified semiclassical theory of Barut and Zanghi is
compatible with the Dirac equation and its proper time parametrization. This
formulation corresponds to a one-charge theory that admits particle and
antiparticle states, so it describes {\it Zitterbewegung} as well as
particle creation processes. The classical theories such as those developed
by Frenkel, Thomas, and Bargmann, Michel, and Telegdi, describe only a
positive-mass system. In other words, to obtain in the classical limit the
BMT equation it is necessary to project the generalized equations of motion
on the positive mass subspace. This is the essential step in order to go
from Beck's proper time derivative to Fock's one.

In summary we have established a deeper connection among classical and
quantum theories of spinning systems, coming from a semiclassical treatment
--which, however, inherits the old problems of the Dirac theory-- to
heuristic quantum proper time derivatives, which acquire meaning in the
proper time parametrization of the Dirac equation. Using this
parametrization we can retrieve, through a semiclassical limit, the
classical covariant equations of motion in proper time. It supports another
evidence of the effectiveness of the Feynman parametrization to deal with
spin systems.\footnote{%
The decisive argument supporting this parametrization is that Feynman \cite
{f48,f51} himself derived QED from it (for an account of these ideas see
Schweber's \cite{s86} review and Refs. \cite{gag97,gag98}).}

\medskip\ \ 

\noindent {\bf Acknowledgments}

\smallskip\ \ 

We are indebted to Bill Schieve, Tomio Petrosky, and Gonzalo Ord\'o\~nez in
occasion of our visit to the University of Texas at Austin, and to Leonid
Burakovsky and Terry Goldman for their hospitality in Los Alamos.


\smallskip 

\begin{thebibliography}{99}
\bibitem{rk63}  {\small S.I. Rubinow and J.B. Keller, {\it Phys. Rev.} {\bf %
131}, 2789 (1963). }\ 

\bibitem{rs64}  {\small K. Rafanelli and R. Schiller, {\it Phys. Rev.} {\bf %
135}, B279 (1964). }

\bibitem{a95a}  {\small J.P. Aparicio, F.H. Gaioli, and E.T. Garcia Alvarez, 
{\it Phys. Rev. A} {\bf 51}, 96 (1995). }

\bibitem{f37}  {\small V. Fock, {\it Phys. Z. Sowjetunion} {\bf 12}, 404
(1937). }

\bibitem{b42}  {\small G. Beck, {\it Rev. Faculdade Ciencias Coimbra} {\bf 10%
}, 66 (1942). }

\bibitem{s4142}  {\small E.C.G. St\"uckelberg, {\it Helv. Phys. Acta} {\bf 14%
}, 322 (1941); {\bf 15}, 23 (1942). }

\bibitem{f48}  {\small R.P. Feynman, {\it Alternative Formulation of Quantum
Electrodynamics}, Pocono Conference, Pennsylvania (1948), quoted in Ref. 
\cite{s86}. }

\bibitem{f50}  {\small R.P. Feynman, {\it Phys. Rev.} {\bf 80}, 440 (1950). }

\bibitem{f51}  {\small R.P. Feynman,{\it Phys. Rev.} {\bf 84}, 108 (1951). }

\bibitem{n50}  {\small Y. Nambu, {\it Prog. Theor. Phys.} {\bf 5}, 82
(1950). }

\bibitem{s51}  {\small J. Schwinger, {\it Phys. Rev.} {\bf 82}, 664 (1951). }

\bibitem{f26}  {\small J. Frenkel, {\it Z. Phys.} {\bf 37}, 243 (1926). }

\bibitem{t27}  {\small L.T. Thomas, {\it Phil. Mag.} {\bf 3}, 1 (1927). }

\bibitem{bmt59}  {\small V. Bargmann, L. Michel, and V.L. Telegdi, {\it %
Phys. Rev. Lett.} {\bf 2}, 435 (1959). }

\bibitem{bz84}  {\small A.O. Barut and N. Zanghi, {\it Phys. Rev. Lett.} 
{\bf 52}, 2009 (1984). }

\bibitem{m37}  {\small M. Mathisson, {\it Acta Phys. Polonica} {\bf 6}, 163
(1937); {\bf 6}, 218 (1937). }

\bibitem{wr47}  {\small J. Weyssenhoff and A. Raabe, {\it Acta Phys. Polonica%
} {\bf 9}, 7 (1947); {\bf 9}, 19 (1947). }

\bibitem{p51}  {\small A. Papapetrou, {\it Proc. Roy. Soc. (London)} {\bf %
A209}, 248 (1951). }

\bibitem{k34}  {\small H.A. Kramers, {\it Physica} {\bf 1}, 825 (1934). }

\bibitem{p32}  {\small W. Pauli, {\it Helv. Phys. Acta} {\bf 5}, 179 (1932). 
}

\bibitem{ga98}  {\small E.T. Garcia Alvarez, {\it PhD Thesis}, Universidad
de Buenos Aires (1998), in preparation. }

\bibitem{mw55}  {\small L. Michel and A.S. Wightman, {\it Phys. Rev.} {\bf 98%
}, 1190 (1955). }

\bibitem{a91}  {\small Aparicio, J.P., Gaioli, F.H., Garcia Alvarez, E.T.,
Hurtado de Mendoza, D.F., and K\'alnay, A.J. (1991). {\it Anales AFA}, {\bf 3%
}, 46. }

\bibitem{a92}  {\small J.P. Aparicio, F.H. Gaioli, and E.T. Garcia Alvarez, 
{\it Anales AFA} {\bf 4}, 9 (1992). }

\bibitem{a95b}  {\small J.P. Aparicio, F.H. Gaioli, and E.T. Garcia Alvarez, 
{\it Phys. Lett. A} {\bf 200}, 233 (1995). }

\bibitem{gag96}  {\small E.T. Garcia Alvarez and F.H. Gaioli, {\it The role
of the de Sitter group in relativistic quantum mechanics}. Proceedings of
the XXI International Colloquium on Group Theoretical Methods in Physics
`Group 21' (1996), Doebner, H.E., Scherer, W., and Natermann, P., eds.,
World Scientific, Singapore, Vol 1, 514. }

\bibitem{gag97}  {\small E.T. Garcia Alvarez and F.H. Gaioli, {\it Int. J.
Theor. Phys.} {\bf 36}, 2391 (1997). }

\bibitem{gag98}  {\small E.T. Garcia Alvarez and F.H. Gaioli, {\it The de
Sitter invariance of QED}, {\it Found. Phys.}, this volume (1998). }

\bibitem{bt85b}  {\small A.O. Barut and W. Thacker, {\it Phys. Rev. D} {\bf %
31}, 2076 (1985). }

\bibitem{f49}  {\small R.P. Feynman, {\it Phys. Rev}}.{\small \ {\bf 76},
749 (1949). }

\bibitem{gga95}  {\small F.H. Gaioli and E.T. Garcia Alvarez, {\it Amer. J.
Phys}}.{\small \ {\bf 63}, 177 (1995). }

\bibitem{p54}  {\small A. Proca, {\it J. Phys. et Rad.} {\bf 15}, 65 (1954). 
}

\bibitem{bp87}  {\small A.O. Barut and M. Pavsic, {\it Class. Quantum Grav.} 
{\bf 4}, L41 (1987); {\bf 4}, L131 (1987). }

\bibitem{bu93}  {\small A.O. Barut and N. Un\"al, {\it Found. Phys}}.{\small %
\ {\bf 23}, 1423 (1993). }

\bibitem{bt85a}  {\small A.O. Barut and W. Thacker, {\it Phys. Rev. D} {\bf %
31}, 1386 (1985). }

\bibitem{h52}  {\small K. Huang, {\it Amer. J. Phys.} {\bf 20}, 479 (1952). }

\bibitem{bb81}  {\small A.O. Barut and A.J. Bracken, {\it Phys. Rev. D} {\bf %
23}, 2454 (1981). }

\bibitem{bc93}  {\small A.O. Barut and M.G. Cruz, {\it J. Phys. A: Math. Gen.%
} {\bf 26}, 6499 (1993). }

\bibitem{s86}  {\small S.S. Schweber, {\it Rev. Mod. Phys.} {\bf 58}, 449
(1986). }
\end{thebibliography}
\end{document}